\documentclass[aps,prd,showpacs,twocolumn,amsmath,10pt,superscriptaddress,floatfix,nofootinbib]{revtex4-1}
\usepackage{amsfonts,amssymb,latexsym,amsmath,epsfig,bm,times}

\newcommand{\be}{\begin{equation}}
\newcommand{\ee}{\end{equation}}
\newcommand{\bea}{\begin{eqnarray}}
\newcommand{\eea}{\end{eqnarray}}
\newcommand{\vep}{{\bm p}}
\newcommand{\veq}{{\bm q}}
\newcommand{\vel}{{\bm l}}
\newcommand{\veep}{{\bm \varepsilon}}

\newcommand{\Tz}{\ensuremath{T_{++,SS}^{(0)}}}
\newcommand{\To}{\ensuremath{T_{++,SS}^{(1)}}}
\newcommand{\Tosd}{\ensuremath{T_{++,SD}^{(1)}}}
\newcommand{\Tods}{\ensuremath{T_{++,DS}^{(1)}}}
\newcommand{\Todd}{\ensuremath{T_{++,DD}^{(1)}}}
\newcommand{\F}{\ensuremath{f_\pi}}

\begin{document}

\title{Remarks on the study of the X(3872) from Effective Field Theory with Pion-Exchange Interaction}

\author{V. Baru}
\affiliation{Institut f\"ur Theoretische Physik II, Ruhr-Universit\"at Bochum, D-44780 Bochum, Germany}
\affiliation{Institute for Theoretical and Experimental Physics, B. Cheremushkinskaya 25, 117218 Moscow, Russia}

\author{E. Epelbaum}
\affiliation{Institut f\"ur Theoretische Physik II, Ruhr-Universit\"at Bochum, D-44780 Bochum, Germany}

\author{A. A. Filin}
\affiliation{Institut f\"ur Theoretische Physik II, Ruhr-Universit\"at Bochum, D-44780 Bochum, Germany} 

\author{F.-K. Guo}
\affiliation{Helmholtz-Institut f\"ur Strahlen- und Kernphysik and
Bethe Center for Theoretical Physics, Universit\"{a}t Bonn, D-53115 Bonn, Germany}

\author{H.-W. Hammer}
\affiliation{Helmholtz-Institut f\"ur Strahlen- und Kernphysik and
Bethe Center for Theoretical Physics, Universit\"{a}t Bonn, D-53115 Bonn, Germany}
\affiliation{Institut f\"ur Kernphysik, Technische Universit\"at Darmstadt, 64289 Darmstadt, Germany}
\affiliation{ExtreMe Matter Institute EMMI, GSI Helmholtzzentrum f\"ur
Schwerionenforschung GmbH, 64291 Darmstadt, Germany}

\author{C. Hanhart}
\affiliation{Forschungszentrum J\"ulich, Institute for Advanced Simulation, Institut f\"ur Kernphysik and
J\"ulich Center for Hadron Physics, D-52425 J\"ulich, Germany}

\author{U.-G. Mei\ss ner}
\affiliation{Forschungszentrum J\"ulich, Institute for Advanced Simulation, Institut f\"ur Kernphysik and
J\"ulich Center for Hadron Physics, D-52425 J\"ulich, Germany}
\affiliation{Helmholtz-Institut f\"ur Strahlen- und Kernphysik and
Bethe Center for Theoretical Physics, Universit\"{a}t Bonn, D-53115 Bonn, Germany}

\author{A. V. Nefediev}
\affiliation{Institute for Theoretical and Experimental Physics, B. Cheremushkinskaya 25, 117218 Moscow, Russia}
\affiliation{National Research Nuclear University MEPhI, 115409, Moscow, Russia}
\affiliation{Moscow Institute of Physics and Technology, 141700, Dolgoprudny, Moscow Region, Russia}


\begin{abstract}
In a recent paper \cite{wangs}, the charmonium state $X(3872)$ is studied in the framework of an effective field theory.
In that work it is claimed that (i) the one-pion exchange (OPE) alone provides sufficient binding to produce the
$X$ as a shallow bound state at the $D^0\bar{D}^{*0}$ threshold, (ii) short-range dynamics (described by a
contact interaction) provides only moderate corrections to the OPE, and (iii) the $X$-pole disappears as the pion
mass is increased slightly and therefore the $X$ should not be seen on the lattice, away from the
pion physical mass point, if it were a molecular state. 
In this paper we demonstrate that the results of Ref.~\cite{wangs} suffer from technical as well as conceptual problems
and therefore do not support the conclusions drawn by the authors.
\end{abstract}

\maketitle

\section{Introduction}

The first evidence for the existence of a narrow ($\Gamma_X<1.2~\mbox{MeV}$ \cite{PDG}) char\-mo\-ni\-um-like state
$X(3872)$ was reported in 2003 by the Belle Collaboration \cite{Xobservation}. The properties of this state
are inconsistent with a simple quark--antiquark meson interpretation, and thus it has
attracted and is still attracting a lot of attention from both
theorists and experimentalists. The $X(3872)$ has the mass~\cite{PDG}
\be
M_X=(3871.68\pm 0.17)~\mbox{MeV}
\label{Xmass}
\ee
and therefore it resides within less than 1~MeV from the neutral $D\bar{D}^*$ threshold. The latter fact implies that
the admixture of the $D^0\bar{D}^{*0}$ component in the wave function of the $X$ can be substantial. Its quantum numbers
are determined to be $J^{PC}=1^{++}$~\cite{LHCb}.

An issue related to the $X(3872)$ as a molecular state heavily discussed in the literature is 
 the nature of the binding forces forming it as a near-threshold
state.
Pion exchange between charmed mesons was suggested long ago \cite{molecule1,molecule2} as a mechanism able to bind
the isosinglet $D \bar D^{*}$ mesonic system and to form a deuteron-like state near threshold. This model was
revisited shortly after the $X(3872)$ discovery \cite{molecule3,molecule4}, while further implications of the nearby
pion threshold are discussed in Refs.~\cite{braatenpions,pions}. 

Because of
the $P$-wave nature of the $D^*D\pi$ coupling,
the OPE potential does not fall off and stays finite at large momenta. 
As such, it contains short-ranged physics
and all loop integrals with the pion exchange are divergent. 
This observation explains the large regulator dependence observed
in Ref.~\cite{ThCl}\footnote{The role of the short-range contribution of the 
pion exchange is also discussed in Ref.~\cite{ch}.}, where calculations were performed in the framework of
a phenomenological potential model with static $D$ mesons.
However, the $D^{*0}$ mass is very close to the $D^0 \pi^0$ threshold and thus 
the intermediate pion may go on-shell~\cite{suzuki}. In consequence, the three-body $D\bar{D}\pi$ unitarity
cuts have to be taken into account\footnote{It is shown in Ref.~\cite{deeply} that cut effects are of paramount
importance
in the $\bar{D}_{\alpha}D_{\beta}$ system, if the $D_{\beta}$
width is dominated by the $S$-wave $D_{\beta}\to
D_{\alpha}\pi$
decay.}. The effects of three-body cuts were included in the effective field theory treatments with perturbative
pions (the X-EFT) \cite{pions}
and nonperturbative pions \cite{Baru:2011rs} based on Faddeev-type integral equations\footnote{In
Ref.~\cite{Kalashnikova:2012qf}
the role of relativistic corrections in the nonperturbative approach including three-body effects was addressed.}.
Both approaches were recently extended to investigate the pion mass dependence
of the $X$ binding energy, see Ref.~\cite{Jansen:2013cba} for the X-EFT
study and  Ref.~\cite{Baru:2013rta} for the nonperturbative calculation.
In these works the behaviour of the $X$  binding
energy was found to be nontrivial:
depending on the interplay of long- and short-range forces
the $X$ can either disappear as a bound state or get more bound. 

In a recent paper \cite{wangs}, the authors revisit the problem of the binding forces in the $X(3872)$ using an
effective
field theory approach. They claim that the OPE alone provides sufficient binding to
produce the $X$ as a shallow bound state at the $D^0\bar{D}^{*0}$ threshold, while the short-range dynamics (described
by a contact interaction) provides only moderate corrections to the OPE. In this paper we demonstrate that the
conclusions drawn by the authors of Ref.~\cite{wangs} are incorrect.
First, the authors apply dimensional regularization to
linearly divergent one-loop integrals and misinterpret their apparent
finiteness for $D=4$, with $D$ being the space-time dimensions, as an ability to
disentangle the OPE and the short-range physics in a model-independent way.
Furthermore, the authors of Ref.~\cite{wangs} resort to a low-order
Pad{\'e} approximation. We do not only argue that this 
method leads to inaccurate results but also demonstrate that
it produces a large number of unphysical singularities within
the assumed range of applicability of the formalism, so that its
predictions, including the emerging $S$-matrix pole position, cannot have any sensible interpretation. 
In addition, we show that in Ref.~\cite{wangs} the three-body singularities
were treated incorrectly.
In short,
we demonstrate that the
results of Ref.~\cite{wangs} are incorrect and as such are devoid of any physical significance.

The structure of the paper is as follows. In Secs. \ref{sec:tree} and \ref{sec:loop} we provide the expressions
for the tree-level and one-loop amplitudes in analogy to those discussed in Ref.~\cite{wangs}. However, instead of
using dimensional regularization applied in Ref.~\cite{wangs}, we stick to a sharp cut-off regularization in order to
make the divergences in the one-loop
contributions explicit. In Sec.~\ref{sec:unitar} we demonstrate
that the entire approach used in Ref.~\cite{wangs} is inconsistent and therefore argue that the results obtained are
unreliable. Our conclusions are summarized in Sec.\ref{sec:conc}.

For convenience we stick to the definitions, conventions, and notations of Ref.~\cite{wangs}. In particular,
the $C$-even combination of the states $1\equiv \bar{D}D^*$ and $2\equiv D\bar{D}^*$ is chosen in the form
\begin{equation}
|X_+\rangle=\frac{1}{\sqrt{2}}(|\bar{D}D^*\rangle+|D\bar{D}^*\rangle),
\label{eq:X}
\end{equation}
and thus the $D\bar{D}^*$ scattering amplitude under consideration is
\begin{equation}
T_{++}=\langle X_+|\hat{T}|X_+\rangle=\frac{1}{2}(T_{11}+T_{12}+T_{21}+T_{22}).
\label{Tpp}
\end{equation}

\section{The tree-level amplitude}
\label{sec:tree}

\begin{figure}[t!]
\begin{center}
\mbox{\epsfxsize=50mm\epsffile{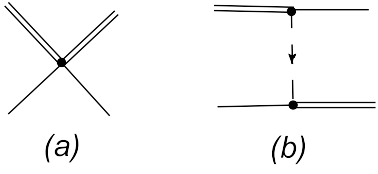}}
\end{center}
\caption{\label{TreeLevel} Tree-level amplitudes. Double, solid, and dashed lines indicate the vector ($D^{*0}$ or
$\bar{D}^{*0}$) mesons, the pseudoscalar ($D^0$ or $\bar{D}^0$) mesons, and the pions, respectively. Adapted from
Ref.~\cite{wangs}.}\label{fig0}
\end{figure}

The general Lagrangian describing four-boson contact interactions is taken as~\cite{Alfiky:06}
\begin{eqnarray}
&&\mathcal{L}^{(0)}=C_2 \Bigl[P^{(Q)\dag} P^{(\bar{Q})} V_{\mu}^{(\bar{Q})\dag} V^{(Q)\mu}\hspace*{2cm}\nonumber\\
&&\hspace*{1.5cm}+ P^{(\bar{Q})\dag} P^{(Q)} V_{\mu}^{(Q)\dag}V^{(\bar{Q})\mu}\Bigr]\nonumber\\
&&\hspace*{2cm}-C_1 \Bigl[ P^{(Q)\dag} P^{(Q)} V_{\mu}^{(\bar{Q})\dag} V^{(\bar{Q})\mu}\\
&&\hspace*{3cm}+ P^{(\bar{Q})\dag} P^{(\bar{Q})}V_{\mu}^{(Q)\dag} V^{(Q)\mu} \Bigr] ,\nonumber
\end{eqnarray}
where ${P}^{(Q)}=(D^0,D^+)$ and $V^{(Q)}=(D^{*0},D^{*+})$ are the heavy meson
fields, while
$\bar{P}^{(\bar Q)}=(\bar{D}^0,D^-)$ and $V^{(\bar{Q})}=(\bar{D}^{*0},D^{*-})$
are the heavy antimeson
fields. The two contact terms $C_1$ and $C_2$ enter the scattering amplitude $T_{++}$ in the combination
$\lambda=C_2-C_1$.

The $D^*D\pi$ interaction relevant for the OPE in the $X(3872)$ is described by the Lagrangian
\begin{eqnarray}
\mathcal{L}^{(1)}&=&2g_{\pi} (V_{a\mu}^{(Q)\dag} P_b^{(Q)} + P_a^{(Q)\dag} V_{b\mu}^{(Q)}) u_{ba}^{\mu}\nonumber\\
&-& 2g_{\pi} (V_{a\mu}^{(\bar{Q})\dag} P_b^{(\bar{Q})} + P_a^{(\bar{Q})\dag} V_{b\mu}^{(\bar{Q})}) u_{ab}^{\mu},
\label{L1}\\
u_{\mu}&=&i(u^{\dag}\partial_{\mu}u-u\partial_{\mu}u^{\dag}),\quad u=\exp\left(\frac{i\phi}{\sqrt{2}\F}\right),
\nonumber
\end{eqnarray}
where $\phi$ is pions matrix, $\F$ is the pion decay constant,
$\F=92.2$~MeV \cite{PDG}. The coupling
constant $g_{\pi}$ is conventionally defined as
\be
g_{\pi}=g\sqrt{M_D M_{D^*}},
\ee
where the dimensionless constant $g$ is determined from the strong decay $D^{*+}\rightarrow D^+\pi^0$. Following
Ref.~\cite{wangs} we use $g=0.3$.

As follows from Lagrangian (\ref{L1}), the $D^*$ decays into $D\pi$ in a $P$ wave, so that the corresponding vertex
contains the pion momentum. Then the components of the tree-level amplitude (see Fig.~\ref{fig0}) are
\cite{wangs,Wangs:unpublished}
\bea
T^{(0)}_{++,SS}&=&-\lambda-\frac{2g_{\pi}^2}{3\F^2}\left[2-\frac{\mu_{\pi}^2}{2\vep^2}\ln\left(1+\frac{4\vep^2}{\mu_{\pi
}^2}
\right)\right],\nonumber\\
T^{(0)}_{++,SD}&=&T^{(0)}_{++,DS}=\frac{2g_\pi^2}{3\sqrt{2}f_\pi^2}\left[
1-\frac{3\mu_\pi^4}{2\vep^2}\right.\nonumber\\
&&\hspace*{6mm}+\left.\left(\frac{\mu_\pi^2}{2\vep^2}+\frac{3\mu_\pi^4}{8\vep^4}\right)\ln\left(1+\frac{
4\vep^2 } { \mu_\pi^2 }\right)\right],\label{T0}\\
T^{(0)}_{++,DD}&=&\frac{g_\pi^2}{3f_\pi^2}\left[
1-\frac{3\mu_\pi^2}{\vep^2}-\frac{9\mu_\pi^4}{4\vep^4}\right.\nonumber\\
&&+\left.\left(\frac{\mu_\pi^2}{2\vep^2}+\frac{15\mu_\pi^4}{8\vep^4}+\frac{9\mu_\pi^6}{16\vep^6}
\right)\ln\left(1+\frac{4\vep^2 } {\mu_\pi^2 }\right)\right],\nonumber
\eea
where $\vep$ is the three-momentum of the $D$-meson in the 
centre-of-mass frame of the $D\bar{D}^*$ system,
\be
\mu_{\pi}^2=m_{\pi}^2-\Delta^2,\quad \Delta=M_{D^*}-M_D,
\label{mupi}
\ee
and the terms $\propto g_\pi^2$ result from the angular integration of the OPE interaction.
To derive Eq.~(\ref{T0}) the terms $\simeq\frac{\displaystyle m_\pi}{\displaystyle M_D}\vep^2$ and $\simeq m_\pi E$ 
($E$ is the $D\bar{D}^*$ energy relative to the two-body threshold) in the three-body propagator
were dropped. As we will argue below in some more detail this significantly changes the singularity structure
of the amplitude. Thus, following the authors of Ref.~\cite{wangs}, 
we consider on-shell $D$ and $D^*$ mesons in their centre-of-mass frame 
with the incoming $D$- and $D^*$-meson momenta
\be
p_1^\mu=\left(E_\vep,\vep\right),\quad p_2^\mu=\left(E^*_\vep,-\vep\right),
\label{cm12}
\ee
and their outgoing momenta
\be
p_3^\mu=\left(E_{\vep'},-\vep'\right),\quad p_4^\mu=\left(E^*_{\vep'},\vep'\right),
\label{cm34}
\ee
with $|\vep'|=|\vep|$. The corresponding $D$- and $D^*$-meson energies are
\be
E_\vep=\sqrt{\vep^2+M_D^2},\quad E^*_\vep=\sqrt{\vep^2+M_{D^*}^2}.
\ee

\section{The one-loop amplitude}
\label{sec:loop}

\begin{figure}[t!]
\begin{center}
\mbox{\epsfxsize=80mm\epsffile{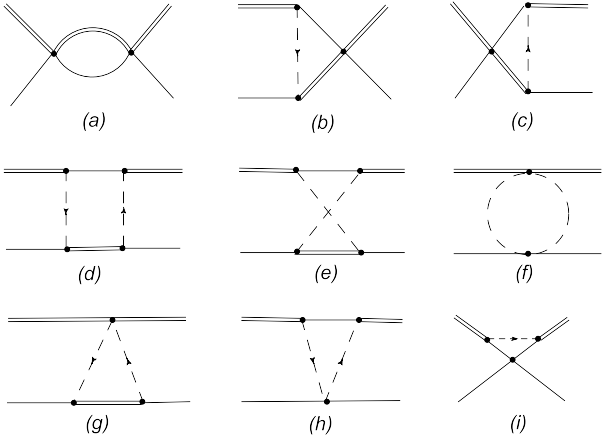}}
\end{center}
\caption{\label{one-loop} One loop diagrams. Adapted from
Ref.~\cite{wangs}.}\label{fig1}
\end{figure}

The diagrams contributing to the one-loop $D\bar{D}^*$ scattering amplitude are depicted in Fig.~\ref{fig1}.
The amplitude for the diagram (a) corresponds to a contact term (pionless) theory and reads
\be
iT^{(1),a}_{++}=\lambda^2\varepsilon_\mu(p_1)\varepsilon_\nu^*(p_3)\int\frac{d^4l}{(2\pi)^4}G(p_1-l)G_{*\mu\nu}(p_2+l),
\label{T1a0}
\ee
where, as was explained before, $p_1$ and $p_2$ are the incoming $D$- and $D^*$-meson momenta, respectively,
$\varepsilon_\mu(p_1)$ and
$\varepsilon_\nu^*(p_3)$ are the initial and final $D^*$ polarization vectors and
\be
G(p)=\frac{1}{p^2-M_D^2+i\varepsilon},\quad
G_{*\mu\nu}(p)=\frac{-g_{\mu\nu}+p_\mu p_\nu/M_{D^*}^2}{p^2-M_{D^*}^2+i\varepsilon}
\ee
are the $D$- and $D^*$-meson propagators, respectively. In the nonrelativistic limit (to the order $O(p)$) only the
spatial Kronecker structure is retained in the $D^*$ propagator, so that the amplitude $T^{(1),a}_{++}$ reads
\be
T^{(1),a}_{++}=(\veep\cdot\veep^*)T^{(1),a}_{++,SS},\quad T^{(1),a}_{++,SS}=\lambda^2I,
\ee
where the loop function $I$ is given by the integral
\bea
I&=&-i\mu^{4-D}\int\frac{d^Dl}{(2\pi)^D}\frac{1}{(p_1-l)^2-M_D^2+i\varepsilon}\nonumber\\
&\times&\frac{1}{(p_2+l)^2-M_{D^*}^2+i\varepsilon},
\label{Idef}
\eea
which is logarithmically divergent in $D=4$. Here $\mu$ stands for a renormalization scale in dimensional
regularization. 
 A straightforward way to deal with this logarithmic divergence is to
employ a convenient regularization procedure (sharp cut-off, dimensional regularization, and so on) and finally to
absorb the divergent piece into the redefinition of the interaction strength $\lambda$ (see, for example,
Ref.~\cite{Alfiky:06} for a contact interaction theory and Ref.~\cite{Baru:2011rs} for the OPE included).

Here, following Ref.~\cite{wangs}, we first perform the integration over the energy $l_0$ by closing the contour in the
upper half of the $l_0$ complex plane. The two relevant poles are
\be
l_0^{(1)}=E_\vep-E_{\vep-\vel}+i\varepsilon,\quad l_0^{(2)}=-E^*_\vep-E^*_{\vep+\vel}+i\varepsilon ~.
\ee
However, anticipating the nonrelativistic expansion we retain only the contribution of the pole $l_0^{(1)}$. This
yields
\be
I=\int\frac{d^{D-1}l}{(2\pi)^{D-1}}\frac{\mu^{4-D}}{2E_{\vep-\vel}[E^{*2}_{\vep-\vel}-(E_\vep+E^*_\vep-E_{
\vep-\vel})^2]}.
\label{Idef2}
\ee
Now, treating the $D$ and $D^*$ mesons nonrelativistically, we expand their energies keeping only the leading
contribution. Then
\be
I=\frac{\mu^{4-D}}{2(M_D+M_{D^*})}\int\frac{d^{D-1}l}{(2\pi)^{D-1}}\frac{1}{(\vel-\vep)^2-\vep^2-i\varepsilon}.
\label{I3}
\ee
Because of the substitution $2E_{\vep-\vel}\approx 2M_D$ made in the denominator, the remaining integral is now linearly
divergent in $D=4$. Finally, using dimensional regularization one arrives at the expression
\be
I=\frac{\mu^{4-D}\Gamma\left(\frac{3-D}{2}\right)}{2(M_D+M_{D^*})(4\pi)^{\frac{D-1}{2}}(-\vep^2-i\varepsilon)^{\frac{3-D
}{2
}}}
\label{Idr3}
\ee
which reveals the well-known feature of dimensional regularization to hide power-like divergences,
in particular, the linear one.
Indeed, formally setting $D=4$ in Eq.~(\ref{Idr3}) one arrives at the finite result
\be
I_{\rm naive}=\frac{i|\vep|}{8\pi(M_D+M_{D^*})}
\label{Inaive}
\ee
used in Ref.~\cite{wangs} --- see Eq.~(12) of Ref.~\cite{wangs}.

It has to be noticed however that the finite expression (\ref{Inaive}) is a result of an implicit subtraction hidden by
the dimensional regularization scheme. To make the argument more transparent we use the sharp cut-off regularization
scheme for the integral (\ref{I3}) to arrive at
\be
I=\frac{1}{8\pi(M_D+M_{D^*})}\left(\frac{2}{\pi}\Lambda+i|\vep|\right),
\label{Ires}
\ee
where $\Lambda$ is the cut-off parameter, so that
\be
T^{(1),a}_{++,SS}=\frac{\lambda^2}{8\pi(M_D+M_{D^*})}\left(\frac{2}{\pi}\Lambda+i|\vep|\right).
\label{T1a}
\ee

In fact, the power-like divergence manifests itself as a pole at $D=3$ in
Eq.~\eqref{Idr3}. If we subtract this divergence as well, following the power
divergence subtraction (PDS) scheme~\cite{Kaplan:1998tg}, we arrive at the same
expression as in Eq.~\eqref{Ires} with $2\Lambda/\pi$ replaced by the scale in
the PDS scheme.

The diagrams (b)-(i) in Fig.~\ref{fig1} contribute to the one-loop amplitude in the pionfull theory. The authors of
Ref.~\cite{wangs} restrict themselves to the order $O(p)$, so that only the contributions from the diagrams (b)-(d) are
retained. For the amplitude in Fig.~\ref{fig1}(b) one has
\begin{widetext}
\be
iT_{++}^{(1),b}=\frac{4\lambda g_\pi^2}{\F^2}\varepsilon_\mu(p_1)\varepsilon_\nu^*(p_3)\int\frac{d^Dl}{(2\pi)^D}
l_\mu l_\lambda G(p_1-l)G_{*\nu\lambda}(p_2+l)D_\pi(l),\quad D_\pi(p)=\frac{1}{p^2-m_\pi^2+i\varepsilon}.
\ee
As before, the integration over the energy $l_0$ is performed explicitly, and the leading nonrelativistic contribution
reads \cite{Wangs:unpublished}
\be
T_{++}^{(1),b}=\frac{2\lambda g_\pi^2\mu^{4-D}}{\F^2(M_D+M_{D^*})}\int\frac{d^{D-1}l}{(2\pi)^{D-1}}
\frac{(\veep\cdot\vel)(\veep^*\cdot\vel)}{[(\vel-\vep)^2-p^2-i\varepsilon][\vel^2+\mu_\pi^2-i\varepsilon]}.
\label{T1b0}
\ee
Note that here the same approximations were made as in the derivation of the tree-level amplitude (see the discussion
after Eq.~(\ref{T0})). 
In order to avoid implicit subtractions we, again, evaluate the linearly divergent integral in Eq.~(\ref{T1b0}) in $D=4$
and
using the sharp cut-off prescription. The result reads
\be
T_{++,SS}^{(1),b}=\frac{\lambda g_\pi^2}{6\pi \F^2(M_D+M_{D^*})}\left(\frac{2}{\pi}\Lambda-
\frac12\mu_\pi^2\Gamma_0(|\vep|)+i|\vep|\right),\quad \Gamma_0(|\vep|)=\frac{1}{|\vep|}\left[
\arctan\frac{2|\vep|}{\mu_\pi}+\frac{i}{2}\ln\left(1+\frac{4\vep^2}{\mu_\pi^2}\right)
\right].
\label{T1b}
\ee
The diagram depicted in Fig.~\ref{fig1}(c) gives the same contribution. Similarly to the amplitude (\ref{T1a}),
using the dimensional regularization scheme in the amplitude (\ref{T1b0}) hides the divergence, so that the result
reported in Ref.~\cite{wangs} corresponds to the divergent term $\propto\Lambda$ in parenthesis implicitly subtracted
---
see Eq.~(13) of Ref.~\cite{wangs}.

Finally, the amplitude for the box diagram depicted in Fig.~\ref{fig1}(d) reads
\be
T_{++}^{(1),d}=\frac{16g_\pi^4}{\F^4}\mu^{4-D}\varepsilon_\mu(p_1)\varepsilon_\nu^*(p_3)\int\frac{d^Dl}{(2\pi)^D}
l_\mu l_\lambda (l+q)_\nu (l+q)_\sigma G(p_1-l)G_{*\nu\lambda}(p_2+l)D_\pi(l+q)D_\pi(l),\quad q=p_3-p_1
\ee
or, after performing the integration over $l_0$ and retaining only the leading contribution \cite{Wangs:unpublished},
\be
T_{++}^{(1),d}=\frac{8g_\pi^4\mu^{4-D}\varepsilon_i\varepsilon_k^*}{\F^4(M_D+M_{D^*})}
\int\frac{d^{D-1}l}{(2\pi)^{D-1}}
\frac{\vel_i(\vel+\veq)_k\vel\cdot(\vel+\veq)}{[(\vel-\vep)^2-\vep^2-i\varepsilon]
[(\vel+\veq)^2+\mu_\pi^2-i\varepsilon][\vel^2+\mu_\pi^2-i\varepsilon]}.
\label{T1d0}
\ee

In $D=4$ with the sharp cut-off regularization this gives
\be
T_{++,SS}^{(1),d}=\frac{4g_\pi^4\Lambda}{3\pi^2 \F^4 (M_D+M_{D^*})}+\left(T_{++,SS}^{(1),d}\right)_{\rm fin},
\label{T1d}
\ee
where the finite part $\left(T_{++,SS}^{(1),d}\right)_{\rm fin}$ is quoted in Eq.~(\ref{ImT1SS}) below. As before, only
this
finite part survives if the dimensional regularization scheme is naively applied to the linearly divergent integral in
Eq.~(\ref{T1d0}), as it is done in Ref.~\cite{wangs}.

Combining the results (\ref{T1a}), (\ref{T1b}), and (\ref{T1d}), one can find for the one-loop amplitude
$T_{++,SS}^{(1)}$ to the order $O(p)$:
\be
T_{++,SS}^{(1)}=\frac{\Lambda}{\pi^2(M_D+M_{D^*})}\left(\frac14\lambda^2+\lambda\frac{2g_\pi^2}{3\F^2}
+\frac{4g_\pi^4}{3\F^4}\right)+\left(T_{++,SS}^{(1)}\right)_{\rm fin},
\label{T1}
\ee
\end{widetext}
where $\left(T_{++,SS}^{(1)}\right)_{\rm fin}$ is the finite amplitude used in
Ref.~\cite{wangs} instead of
the full one-loop amplitude (\ref{T1}). Notice that the divergent piece $\propto\Lambda$ contains contributions both
from the contact interaction as well as from the OPE. In order to renormalize the one-loop amplitude
(\ref{T1}) the constant contact counterterm has to be added to $T_{++,SS}^{(1)}$, the divergent part of which,
\be
\delta T_{++,SS}^{(1)}=-\frac{\Lambda}{\pi^2(M_D+M_{D^*})}\left(\frac14\lambda^2+\lambda\frac{2g_\pi^2}{3\F^2}
+\frac{4g_\pi^4}{3\F^4}\right),
\label{dT1}
\ee
does not vanish even in the limit $\lambda=0$,
\be
\delta
T_{++,SS}^{(1)}(\lambda=0)=-\frac{4g_\pi^4\Lambda}{3\pi^2\F^4(M_D+M_{D^*})}.
\label{dT11}
\ee
Thus, contrary to the claim of Ref.~\cite{wangs}, setting $\lambda=0$ does not imply
that only the OPE interaction is left, since the contact operator (\ref{dT11}) is added to the OPE.
As was explained
before, this contact operator is hidden (added implicitly) in Ref.~\cite{wangs} by using the
dimensional regularization scheme for linearly divergent integrals. Alternatively to the sharp cut-off scheme used
above,
one can resort to the PDS scheme~\cite{Kaplan:1998tg} and subtract the power divergence in $D=3$. This would reveal the
divergence and make the subtraction explicit. Furthermore, had the authors of
Ref.~\cite{wangs} proceeded beyond the one-loop approximation, divergences would have become explicit, too.

Therefore, the conclusion one is led to is
that the OPE potential in the $D\bar{D}^*$ system is well defined in the sense of an effective field theory
only in connection with a contact operator. Thus the
conclusion drawn in Ref.~\cite{wangs} that ``the pion exchange interaction is the main reason for the system to
be bound'' has to be considered as model- and scheme-dependent.

\section{Pad{\'e} approximation and the would-be $X$ pole}
\label{sec:unitar}

In addition to the conceptual problem outlined in the previous section, the work
of Ref.~\cite{wangs} also suffers from a severe technical problem as we
explain in this section. Based on a particular method of unitarization, 
Ref.~\cite{wangs} reports the existence of a dynamically generated $S$-matrix pole which is
interpreted as the $X(3872)$. In particular, for $\lambda=0$, the pole resides at
\be
p_0=-15.46+i24.62~\mbox{MeV}.
\label{p0}
\ee
In this section we provide very strong evidence that this pole is an artifact 
most probably 
caused by using Pad{\'e}
approximants of a too low order as well as an incorrect treatment of three-body effects.
In particular, to formulate our argument we focus on a regime where the pion is sufficiently heavy that the $D^*$ 
is stable which implies $\mu_\pi^2>0$ (see Eq.~(\ref{mupi})). In this
regime no three-body cuts need to be considered when solving the equations. In addition,
the $S$-matrix needs to be consistent with the general theorems on two-body scattering.
We show that in this regime the $S$-matrix of Ref.~\cite{wangs} contains many singularities most of those being
unphysical and very close to the threshold. This suggests that the formalism used
in Ref.~\cite{wangs} should not be used in the regime $\mu_\pi^2<0$ either. This statement
is further supported by the observation that the three-body effects are treated inconsistently 
in Ref.~\cite{wangs}.

Using Pad{\'e} approximation is a well known technique to solve integral equations in the context of few-body problems
 (see, for example, the very pedagogical presentation in chapter 2.7.3 of the textbook Ref.~\cite{gloecklebook}). 
In general it is argued that one can approximate the physical amplitude $f(E,\xi)$ --- represented
via an infinite series with an increasing number of insertions of the scattering potential scaled by the
strength parameter $\xi$ (where $\xi=1$ refers to the physical situation) ---
by the rational function $f_{[N,M]}(E,\xi)=P_N(\xi)/Q_M(\xi)$, where $P_N(\xi)$ and $Q_M(\xi)$
are polynomials in $\xi$ of order $N$ and $M$, respectively, with energy-dependent coefficients.
In particular, for the case of $NN$ scattering in the spin-singlet $S$-wave channel at 12 MeV above threshold 
(see Tab. 2.1 in Ref.~\cite{gloecklebook}) the Pad{\'e} series 
fully converges only after the inclusion of 10 terms (which implies N=5, M=4). Using only 5 iterations as
input (N=2, M=2) the series is still off by 50\%. The absolute 
value of the
real part of the amplitude is too large by a factor 50 at $N=1$ and $M=0$.
In the three-body case, the situation is similar (see Tab.~3.2 
in Ref.~\cite{gloecklebook} for instance). Naturally, the order of the Pad{\'e} approximation needed to gain an
acceptable accuracy is increased dramatically in case when near-threshold singularities are present in the amplitude. 
It should be stressed at this point that the formalism used in Ref.~\cite{wangs} pretends to be able to describe
a near-threshold bound-state pole and nevertheless it refers to $N=0$ and $M=1$. 
In addition to the
low accuracy that one should expect in this case we will demonstrate below that
for these small values of $N$ and $M$ the amplitude contains several unphysical singularities
within the assumed range of applicability of the formalism.

For the physical pion mass, the
parameter $\mu_\pi$ is purely imaginary
\be
\mu_\pi=-i\sqrt{\Delta^2-m_\pi^2} = -i44.36~\mbox{MeV};
\label{mupiim}
\ee
the $D\bar{D}^*$-threshold as well as the mass of the $X(3872)$ are located
above the $D\bar D\pi$ threshold. A pole search in this kinematic regime
is technically quite demanding. We therefore study the pole structure of
the equations for larger pion masses with $\mu_\pi^2>0$. This simplifies the analysis
significantly and still allows us to show that the equations used in Ref.~\cite{wangs}
are ill-behaved. For real values of $\mu_\pi$, the $D^*$ does not have phase space to decay into $D\pi$, so that
three-body ($D\bar D \pi$) intermediate states can not go on-shell. For simplicity, in this section we use $\lambda=0$.

The equation underlying the results of Ref.~\cite{wangs} are the [0,1] Pad{\'e}
approximation (Eq.~(15) of Ref.~\cite{wangs}) given by
\be
T_{++}^{\rm
phy}=T_{++}^{(0)}\cdot[T_{++}^{(0)}-T_{++}^{(1)}]^{-1}\cdot T_{++}^{(0)},
\label{Tph}
\ee
where the tree-level amplitude $\Tz$ is quoted in Eq.~(\ref{T0}) while, for
$\lambda=0$, the one-loop amplitude $T^{(1)}$ is given entirely by the box
diagram (d) in Fig.~\ref{one-loop} and reads~\cite{Wangs:unpublished}
\begin{widetext}
\bea
\To&=&\frac{g_\pi^4}{12\pi f_\pi^4(M_D+M_{D^*})}\left\{
-4\mu_\pi-\frac{\mu_\pi^3}{\vep^2}-\left(4|\vep|+\frac{6\mu_\pi^2}{|\vep|}\right)\arctan\frac{2|\vep|}{\mu_\pi}+
\left(\frac{\mu_\pi^5}{2\vep^4}+\frac{3\mu_\pi^7}{8\vep^6}\right)\ln\left(1+\frac{\vep^2}{\mu_\pi^2}
\right)\right.\nonumber\\
&+&\left(4|\vep|+\frac{4\mu_\pi^2}{|\vep|}-\frac{\mu_\pi^4}{2|\vep|^3}-\frac{3\mu_\pi^6}{4|\vep|^5}\right)\arctan\left(
\frac{\mu_\pi |\vep|}{\mu_\pi^2+2\vep^2}\right)\label{ImT1SS}\\
&+&\left(\frac{\mu_\pi^2}{|\vep|^3}+\frac{\mu_\pi^6}{2|\vep|^5}+\frac{3\mu_\pi^8}{16|\vep|^7}\right)
\left[\mbox{ImLi}_2\left(\frac{2\vep^2-i|\vep|\mu_\pi}{\mu_\pi^2+4\vep^2}\right)+\mbox{ImLi}_2\left(\frac{
-2\vep^2+i|\vep|\mu_\pi}{
\mu_\pi^2+4\vep^2 }\right)\right]\nonumber\\
&+&\left. i|\vep|\left[3-\frac{\mu_\pi^2}{\vep^2}+\frac{3\mu_\pi^4}{4\vep^4}-\left(\frac{\mu_\pi^2}{\vep^2}+
\frac{\mu_\pi^4}{4\vep^4}+\frac{3\mu_\pi^6}{8\vep^6}\right)\ln\left(1+\frac{4\vep^2}{\mu_\pi^2}\right)
+
\left(\frac{\mu_\pi^4}{4\vep^4}+\frac{\mu_\pi^6}{8\vep^6}+\frac{3\mu_\pi^8}{64\vep^8}\right)\ln^2\left(1+\frac{4\vep^2}{
\mu_\pi^2} \right)
\right]
\right\},\nonumber
\eea
where Li$_2$ is the dilogarithm function\footnote{Notice that in
Eq.~(\ref{ImT1SS}) the divergent piece is subtracted, as was explained in Sect.~\ref{sec:loop}
above.}. The other components,
$\Tosd=\Tods$ and $\Todd$, take a complicated form,
similar to Eq.~(\ref{ImT1SS}) \cite{Wangs:unpublished}, and we do not quote them here.
\end{widetext}

For real values of $\mu_\pi$ one easily verifies that the tree-level and the one-loop
amplitudes satisfy the perturbative two-body unitarity condition (Eq.~(14) of Ref.~\cite{wangs})
\be
\mathrm{Im}T^{(1)}_{++}=T_{++}^{(0)}\frac{|\vep|}{8\pi\sqrt{s}}T^{(0)*}_{++},
\label{ImT1}
\ee
which, for $T^{(1)}_{++,SS}$, takes the form
\be
\mbox{Im}T^{(1)}_{++,SS}=\frac{|\vep|}{8\pi\sqrt{s}}\left(\left|T^{(0)}_{++,SS}\right|^2+
\left|T^{(0)}_{++,SD}\right|^2\right).
\label{ImT1new}
\ee
With this input $T_{++}^{\rm phy}$ defined in Eq.~(\ref{Tph}) is consistent with two-body unitarity.

The authors of Ref.~\cite{wangs} claim that they do not find any pole of the $S$-matrix for $\mu_\pi^2>0$.
In particular, they claim that ``When $m_\pi$ is larger than $\Delta$ (142 MeV), there is no bound
state or resonance pole.'' Meanwhile, by an explicit calculation one can demonstrate that the equation
\be
\mbox{det}\left(T_{++}^{(0)}-T_{++}^{(1)}\right)=0
\ee
for real $\mu_\pi$ does possess multiple solutions similar to the solution (\ref{p0}). For example, for
$\mu_\pi=44.36$~MeV, the following near-threshold solutions exist:
\bea
&&\pm 11.80+i22.47~\mbox{MeV},\quad\pm 23.87-i20.44~\mbox{MeV},\nonumber\\
&&\pm 29.94+i13.24~\mbox{MeV},\quad \pm 10.74-i0.03~\mbox{MeV},\label{sols}\\
&&\pm 10.66+i0.02~\mbox{MeV}.\nonumber
\eea
Each solution above corresponds to a pole of the physical amplitude (\ref{Tph}). In particular, the pole
\be
p_1=-11.80+i22.47~\mbox{MeV}
\label{p1}
\ee
looks very similar to the pole (\ref{p0}) reported in Ref.~\cite{wangs} and, naively, leads to a similar interpretation
as a bound state. It is easy to see, however, that such an interpretation is misleading. Indeed, if the
three-body threshold is not open, the bound state pole in
the complex momentum plane must reside on the positive half of
the imaginary axis (this corresponds to a pole below threshold on the
real axis on the first Riemann sheet of the complex energy plane) in accordance
with general principles of Quantum Mechanics. The fact that $p_1$ from
Eq.~(\ref{p1}) possesses a real part (and quite a large one!) can only be ascribed to a shortcoming of the
low-order Pad{\'e} approximation used in the calculations to produce spurious poles which,
as a matter of principle, cannot be
interpreted as observable objects. The latter statement can be given additional strong support based on the
following argument: by changing the sign
of the OPE, and thus making it repulsive, one would expect all physical poles to go away from the near-threshold
region. Meanwhile, solutions similar to those from Eq.~(\ref{sols}) continue to exist.

In addition to the issues already mentioned, the equations of Ref.~\cite{wangs} suffer from an incorrect treatment
of three-body effects. First of all, in three-body systems there is typically a subtle cancellation
between the imaginary parts that come from the single particle ($D^*$) self-energies embedded in the 
three-body system and those that come from the three-particle ($D\bar D\pi$) propagators~\cite{AAY}.
In Ref.~\cite{wangs} the cuts related to the $D^*$ self-energy are omitted altogether. 
Thus, the imaginary parts from the three-body effects are calculated in an inconsistent manner and are therefore
incorrect. 

In addition, since the recoil terms are dropped in the three-body propagators,
the $D\bar D\pi$ three-body cut is effectively converted into a two-body cut --- notice the singularity at 
$\vep^2=-\mu_\pi^2/4$ in the tree-level amplitude (\ref{T0}) as well as in  
the one-loop amplitude (\ref{ImT1SS}) (for a different reaction
this is discussed in some detail in Ref.~\cite{lensky}). This pronounces the $D\bar D\pi$ singularity
in the very near threshold regime way
too strongly, since a two-body cut scales as $\sqrt{E}$ while
a three-body singularity scales as $E^2$. This again shows 
that the treatment of the three-body dynamics is not correct in Ref.~\cite{wangs}.

\section{Conclusions}
\label{sec:conc}

In this paper we demonstrated that the results of Ref.~\cite{wangs} suffer from technical as well as conceptual problems
and therefore do not support the conclusions drawn by the authors.  We pointed out several flaws of that paper.

First, we demonstrate that the separation of the short-ranged physics
and the OPE which the authors of Ref.~\cite{wangs} dwell on at some length is
not possible as a matter of principle. Thus the conclusions made in
Ref.~\cite{wangs} that the OPE alone provides sufficient binding to produce the
$X$ as a shallow bound state at the $D^0\bar{D}^{*0}$ threshold and that the
short-range dynamics (described by a contact interaction) provides only moderate
corrections to the OPE are incorrect. As proven above,
the OPE potential in the $D\bar{D}^*$ system is well defined in the sense of an
effective field theory only in connection with a contact operator.

Second, based on results found for the two- and three-nucleon system,
we argued that the low-order Pad{\'e} approximation employed
in Ref.~\cite{wangs} to construct the physical amplitude cannot be expected
to be reliable. 

Third, the equations were shown to
produce a significant number of unphysical singularities very close to
the threshold at least 
in the kinematic regime where the $D^*$ is stable. We argue that 
the same should also happen in the physical regime where the $D^*$
is unstable and that there is no reason to expect that the pole reported 
in Ref.~\cite{wangs} is physical.

Fourth, we argued that in Ref.~\cite{wangs} the three-body effects
are treated incorrectly and inconsistently.

Therefore in this paper we have demonstrated that the results of Ref.~\cite{wangs}
are based on an inconsistent and incomplete set of equations that produces
a large number of unphysical singularities in the claimed range of applicability
of the formalism. 
Instead of the formalism of Ref.~\cite{wangs} it appears more 
appropriate to solve the full scattering equation including the one-pion
exchange either perturbatively~\cite{pions} or non-perturbatively~\cite{Baru:2011rs}.
Not only do these formalisms not suffer from the mentioned inconsistencies,
they also allow for a systematic, controlled study of the light-quark mass
dependence of the $X(3872)$. While in Ref.~\cite{wangs} the $X(3872)$ is claimed
to disappear to the second sheet unavoidably as soon as the light-quark
masses are increased slightly, in both Ref.~\cite{Jansen:2013cba}
as well as Ref.~\cite{Baru:2013rta} for perturbative and non-perturbative pions,
respectively, it is demonstrated that field theoretic consistency demands the
inclusion of a quark-mass-dependent contact term in the theory and it is
the sign of that contact term that decides on the fate for the $X(3872)$ pole
as the pion mass is increased. As a consequence, both
Ref.~\cite{Jansen:2013cba}
as well as Ref.~\cite{Baru:2013rta} are consistent with the so
far only existing lattice calculation for the $X(3872)$~\cite{lattice},
while Ref.~\cite{wangs} is not. 
 It should
be stressed that these findings of Refs.~\cite{Jansen:2013cba,Baru:2013rta} for the $D\bar{D}^*$ system are in full
analogy
to those for the nucleon--nucleon
system~\cite{chiralextrapol_seattle,chiralextrapol_bonn,BBN}.

To summarize, we have demonstrated that none of the claims of Ref.~\cite{wangs}
listed in the introduction holds. Especially, under the assumption that the $X(3872)$ is a
$D\bar{D}^*$ molecular state, the pion mass dependence of its pole position is
expected to depend strongly on the pion mass dependence of the $D\bar{D}^*$
interaction at short range. Furthermore, a more deeply bound $X$-state for
increased pion masses as found in Ref.~\cite{lattice} does not contradict  a
molecular nature of the $X$. One might hope to eventually reveal important
information on the structure of the potential responsible for the binding of the
$X(3872)$, once its pion mass dependence is mapped out using lattice QCD from
calculations along the lines of Ref.~\cite{BBN}.

\begin{acknowledgments}
The authors are grateful to P. Wang and X.-G. Wang for
many communications and for providing us with additional, unpublished materials.
We also acknowledge valuable discussions with A. Nogga.
This work is supported by the DFG and the NSFC through funds provided
to the Sino-German CRC 110 ``Symmetries and the Emergence of Structure in
QCD'' (NSFC Grant No. 11261130311), by the NSFC (Grant No. 11165005), by the EU
Integrated Infrastructure Initiative HadronPhysics3 (Grant No. 283286), by the Helmholtz 
Association under contract HA216/EMMI, by the
ERC (259218 NUCLEAREFT), and by NSh-3830.2014.2.
\end{acknowledgments}

\end{document}